\documentclass[showpacs,twocolumn,pre,floatfix]{revtex4}
\usepackage[]{graphicx}
\usepackage[]{color}
\usepackage{amsmath}
\usepackage{amssymb}
\usepackage{xcomment}
\usepackage{ulem}

\newcommand{\rb}{{\bf r}}
\newcommand{\vb}{{\bf v}}

\newcommand{\vct}[1]{\mathbf{#1}}

\begin{document}
\title{
Driven colloidal suspensions in confinement and density functional theory: Microstructure and wall-slip
}
\date{\today}

\author{Artem A. Aerov}
\affiliation{4th Institute for Theoretical Physics, Universit\"at Stuttgart, Germany and Max Planck Institute for Intelligent Systems, 70569 Stuttgart, Germany}
\email{}
\author{Matthias Kr\"uger}
\affiliation{4th Institute for Theoretical Physics, Universit\"at Stuttgart, Germany and Max Planck Institute for Intelligent Systems, 70569 Stuttgart, Germany}
\email{aerov@is.mpg.de}
\begin{abstract}
We theoretically investigate general properties of driven (sheared) colloidal suspensions in confinement, based on methods of classical density functional theory. 
In the absence of an exact closed (Smoluchowski-) equation for the one-particle density under shear, we formulate a set of general conditions for approximations, 
and show that a simple closure fulfills them. The exact microscopic stress tensor is identified. Exemplifying the situation near a wall (oriented parallel to the direction of shear), 
we note that the microscopic shear stress is not necessarily homogeneous. Formulating a second equation additional to the Smoluchowski equation, we achieve a homogeneous shear stress, and thereby compute the 
local flow velocity near the wall. This finally leads to a slip length of the complex fluid at the wall. 
\end{abstract}

\pacs{82.70.Dd, 83.80.Hj, 05.70.Ln
}
\keywords{density functional, diffusion, nonequilibrium}

\maketitle

\section{Introduction}\label{Introduction}
Suspensions of Brownian particles possess a variety of nontrivial properties \cite{dhont,Larson}, especially in the dense regime. Particle interactions result in a slowing down of the internal relaxation
dynamics with increasing density, which eventually turns untractably slow (or glassy) around a packing fraction of roughly 58 \% \cite{Pusey86}. Suspensions are also of great rheological interest \cite{Larson}, 
as particle interactions play a nontrivial role here, too, and lead, e.g., to measurable changes of the suspension's properties in the directions perpendicular to driving. Despite available powerful simulation 
techniques \cite{Allen,Foss00,Berthier02prl,Zausch08,kruger2011}, there are still strong ongoing research activities, aming at describing properties of driven suspensions from first principles. The mentioned 
interplays between flow and structure mediated by particle interactions make microscopic theoretical descriptions desirable. Based on the many particle Smoluchowski equation \cite{Risken}, powerful theoretical 
techniques have been developed for the description of rheological properties of dense suspensions based on Mode Coupling Theory (MCT) \cite{Fuchs02,Miyazaki02,pnas,Fuchs09}. These use density correlation functions 
in (spatial) Fourier space, thereby profiting from the translational invariance of bulk systems. Extensions to confined systems are tedious, but have nevertheless recently been achieved for equilibrium 
situations \cite{Lang10}. Somewhat more phenomenological are the soft-glassy-rheology model \cite{Sollich97} or the shear-transformation-zone model \cite{Falk98}.

The bulk rheology of suspensions is generally characterized by a shear stress, determining the force needed to shear the system. It is naturally linear  in shear rate for small rates, thus defining the linear
response viscosity of the suspension. The diagonal parts of the stress tensor, i.e., the (anisotropic) pressure, change also with applied shear, the deviation from the equilibrium values being quadratic in shear
rate for small rates \cite{Bergenholtz02,brady_morris,Henrich07}, as demanded by symmetry.  

Suspensions in {\it confinement}, their microstructure and phase transitions are in many cases well described {\it in equilibrium} by classical density functional theory (DFT) 
\cite{bob_advances,bob_review, roth_review}. Nonequilibrium situations, such as driven suspensions or suspensions subject to time dependent potentials, have been addressed by Dynamical Density Functional Theory 
(DDFT) \cite{Marconi99,archer, rauscher2}. While the framework described in these references suffices for example to describe the flow disturbance induced by a moving particle \cite{penna, rauscher2}, it does not
capture the effects of a simple shear flow on the microstructure. This shortcoming has been addressed recently \cite{Brader_Kruger_2011,Brader_Kruger_2011_2}, by addition of an extra term. Using this theory, it 
was found that the density can undergo a layering transition into a state with long range order \cite{Brader_Kruger_2011}, and that a colloidal sediment rises under application of shear flow, an effect called 
viscous resuspension. Laning of oppositely charged particles has also been studied by DDFT with addition of an extra term \cite{Chakrabarti03}.  Also, by use of a tagged particle method, DDFT was used
successfully to analyze bulk properties under shear \cite{Johannes_Joe_2013}, and the pair correlation of suspensions with individual driven particles  was investigated on basis of Kirkwood's superposition
approximation \cite{Kohl12}. Recently, a variational principle (``power functional'') has been proposed for study of driven suspensions \cite{Schmidt13}, as well as a general nonequilibrium Ornstein-Zernike
relation \cite{Brader13}. Experimentally, finite (or small) driven systems are readily accessible, see e.g. \cite{Koeppl06,Isa09,Cheng11}. See also a recent review on confined crystals including shear
\cite{Reinmueller13}.   

In this paper, we investigate general properties of sheared systems under confinement, and analyze a description of such systems by use of density functional theory. Specifically, in Sec.~\ref{sec:sys}, we 
introduce the system under consideration together with exact equations describing it. This is followed by a review of the DDFT approximation of Refs.~\cite{Marconi99,archer, rauscher2} in Sec.~\ref{DDFT}. 
In Sec.~\ref{Shear-DDFT}, we introduce the Shear-DDFT approximation, where we first recall the inability of the DDFT approximation of Sec.~\ref{DDFT} to capture effects of shear. We formulate general conditions 
for a Shear-DDFT approximation, and then introduce a simple version. In Sec.~\ref{stress_tensor}, the stress tensor in the exact many body Smoluchowski equation is identified and computed by use of the Shear-DDFT
 approximation. In Sec.~\ref{pressure}, we analyze the diagonal elements of the stress tensor, and show that e.g. the osmotic pressure at the wall is consistent with the bulk pressure in the Shear-DDFT 
approximation. In Sec.~\ref{ensembles}, we note that the off-diagonal part of the stress tensor (the shear stress) is not necessarily homogeneous, and we introduce the {\it stress ensemble} that requires such
 homogeneity. Finally in Sec.~\ref{equations}, we explicitly give results for a hard sphere suspension near a hard wall, computing the density distribution near the wall, the flow velocity in the stress ensemble,
 and the slip length at the wall.
\section{The system}\label{sec:sys}
\subsection{General relations} 
Consider a system of $N$ spherically symmetric Brownian  particles, $i=1\dots N$, immersed in a solvent (compare also Fig.~\ref{fig:1} below). 
Due to the overdamped nature of the dynamics, the distribution of particle positions evolves according to the Smoluchowski equation \cite{archer, dhont}, 
\begin{eqnarray}
\frac{\partial \Psi(t)}{\partial t} = \sum_{i} \nabla_i\cdot \left[D\left(\nabla_i- \beta \,{\bf F}_i\right) - {\bf v}({\bf r}_i) \right]\Psi(t) ,
\label{smol_hydro}
\end{eqnarray} 
where $\Psi(t)\equiv\Psi(\{ {\bf r}_i\},t)$ is the probability function of particle positions, $\beta=1/k_BT$ is inverse thermal energy, and $D$ is the bare diffusion coefficient of the particles;
 ${\bf v}({\bf r}_i)$ is the (driven) flow velocity at the position of particle $i$, ${\bf r}_i$, and ${\bf F}_i$ is the force exerted on particle $i$ by both the external
potential field, and the other particles (with total interparticle interaction energy $U$),
\begin{align}\label{Force}
{\bf F}_i=
-\nabla_i \left[U\{{\bf r}_i\}+ V_{ext}(\rb_i)\right].
\end{align}
We note that Eq.~\eqref{smol_hydro} neglects hydrodynamic interactions, see Sec.~\ref{sec:hyd}.
By integrating Eq.~(\ref{smol_hydro}) over the coordinates of all particles except one, one obtains the equation for the evolution of the one body density $\rho(\rb,t)$ \cite{HansenMcDonald}, which, even more 
obvious than Eq.~\eqref{smol_hydro}, is a continuity equation,
\begin{equation}
\frac{\partial \rho(\rb,t)}{\partial t}=-\nabla \cdot  {\bf j}(\rb,t) \, .
\label{ddft_eq}
\end{equation}
For the case of pair potentials, i.e., $U\{{\bf r}_i\}=\sum_{i,j>i}\phi(|\rb_i-\rb_j|)$, the one body probability current ${\bf j}$ is given by 
\begin{multline}
{\bf j}(\rb,t)=\rho(\rb,t)\vb(\rb)- D \Bigl(\nabla \rho(\rb,t)  \\
+ \beta \rho(\rb,t) \nabla V_{ext}(\rb)+\beta\int\!d\rb'\, \rho^{(2)}(\rb,\rb',t) \nabla_{\rb}\phi(|\rb\!-\!\rb'|)\Bigl).
\label{ddft_current_0}
\end{multline}
Here, $\rho^{(2)}(\rb,\rb',t)$ is the {\it two-particle density} \cite{HansenMcDonald} related to the probability that each of the positions $\rb$ and $\rb'$ is occupied by a particle. 
Despite being exact, Eq.~\eqref{ddft_current_0} is not closed because $\rho^{(2)}$ is unknown in general. 

In the case of {\it thermal equilibrium}, the sum rule is valid \cite{bob_advances, archer}, which introduces the density functional $\mathcal{F}_{\rm ex}$,
\begin{equation}\label{sum_rule}
\int\!d\rb'\, \rho_{eq}^{(2)}(\rb,\rb')\nabla_\rb \phi(|\rb\!-\!\rb'|) = \rho(\rb)\nabla_\rb \frac{\delta \mathcal{F}_{\rm ex}}{\delta \rho(\rb)} \,.
\end{equation}
Substituting (\ref{sum_rule}) into (\ref{ddft_current_0}), and considering the situation without driving, i.e., $\vb(\rb)=0$, the density in Eqs.~\eqref{ddft_eq} and \eqref{ddft_current_0} converges to the
equilibrium density, which minimizes the grand potential $\Omega$ \cite{HansenMcDonald},
\begin{eqnarray}
\Omega[\,\rho\,]=\mathcal{F}_{\rm id}[\,\rho\,] + \mathcal{F}_{\rm ex}[\,\rho\,] + 
\int \!d\rb\, (V^{\rm ext}(\rb) - \mu)\rho(\rb). 
\end{eqnarray}
Here, $\mu$ is the chemical potential of the particles, and it is chosen such that the average particle number $\langle N\rangle$ of the grand canonical ensemble equals the particle number $N$ of
Eq.~\eqref{smol_hydro}. The ideal gas part of the free energy is defined by, 
\begin{eqnarray}
\mathcal{F}_{\rm id}[\,\rho\,]=k_BT\int\!d\rb \rho(\rb)[\ln(\Lambda^3\rho(\rb))-1], 
\end{eqnarray}
where $\Lambda$ is the thermal wavelength. $\mathcal{F}_{\rm ex}[\,\rho\,]$ introduced in Eq.~\eqref{sum_rule} is termed `excess` free energy. It is the contribution of the particle interactions to the free energy
$\mathcal{F}\equiv\mathcal{F}_{\rm id}+\mathcal{F}_{\rm ex}$.  
\subsection{A Comment on hydrodynamic interactions}\label{sec:hyd}
Eq.~\eqref{smol_hydro} results from integrating out the degrees of freedom of the solvent molecules, which results in the stochastic dynamics underlying it. The integration also leads to effective interactions 
between the Brownian particles mediated by the solvent, the so called hydrodynamic interactions \cite{dhont}. Due to their nontrivial nature, they are often neglected in theoretical treatments of 
suspensions, as done in the present work. We are aware that, when introducing the stress ensemble (see Sec. \ref{ensembles} below), a principle interplay between the solvent and the particle stress may be present,
such that treating the two independently, as done in Eq.~\eqref{velocity_eq_general}, may be a crude approximation. Yet, aiming to understand and develop the Shear-DDFT approach from first principles and to keep 
track of the origin of the different effects, we leave the inclusion of hydrodynamic interactions for later work. 

\section{DDFT approximation}\label{DDFT}
Eq.~\eqref{sum_rule} is by definition valid in equilibrium, and in general invalid for systems under driving. Considering the latter, the simplest closure for Eq.~\eqref{ddft_current_0} is achieved by
(now approximative) use of Eq.~\eqref{sum_rule}. Substituting (\ref{sum_rule}) into (\ref{ddft_current_0}), one obtains a closed equation for a system with external driving $\vb(\rb)$ \cite{rauscher2}, 
\begin{multline}
{\bf j}(\rb,t)\approx\rho(\rb,t)\vb(\rb)\\
-\beta D \rho(\rb,t)\nabla \left (\frac{\delta \mathcal{F}[\rho(\rb,t)]}{\delta \rho(\rb,t)} +V_{ext}(\rb)\right).
\label{ddft_current}
\end{multline}
This equation captures the density distortions due to a moving tracer particle \cite{penna, rauscher2}. The use of Eq.~\eqref{sum_rule} in non-equilibrium situations (leading to Eq.~\eqref{ddft_current}) is 
sometimes referred to as adiabatic approximation, as it is based on the assumption that the two-particle function $\rho^{(2)}(\rb,\rb',t)$ relaxes instantaneously in each moment to the one of an equilibrium
system with density profile $\rho(\rb,t)$. 

For the case of simple shear, it is precisely this assumption that will prove too crude to be able to capture the changes in structure, as it is precisely the {\it distortion} of $\rho^{(2)}(\rb,\rb',t)$ due
to shear which contains the coupling of density and flow. 

\section{Shear-DDFT approximation}\label{Shear-DDFT}
\subsection{Repetition: Why Eq.~\eqref{ddft_current} is inappropriate for shear}
\begin{figure*}
\includegraphics[width=0.7\textwidth]{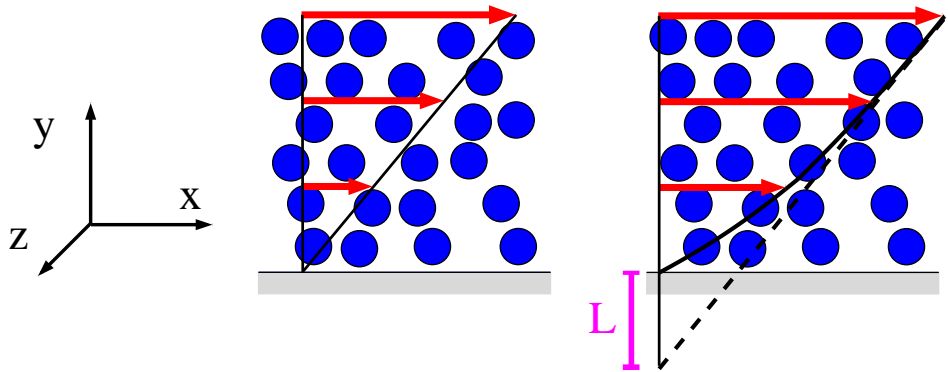}
\caption{\label{fig:1}Specific example configuration for the relations derived in the main text: A suspension of colloidal particles is confined by a wall and subject to an external shear flow. Left hand side
depicts the shear ensemble (see Sec. \ref{ensembles}) where the solvent flow, illustrated by the red arrows, is prescribed. Right hand side depicts the stress ensemble, where the solvent velocity is an output. 
The deviation from the simple shear profile can be quantified by the slip length $L$. }
\end{figure*}

Shear flow  is a simple but yet important example for external driving, as it is the basis for  studying the dynamic viscosity of complex fluids. As discussed in detail in
Refs.~\cite{Brader_Kruger_2011,Brader_Kruger_2011_2}, the scenario depicted in Fig.~\ref{fig:1} cannot  be studied by use of Eq.~\eqref{ddft_current}; The external flow ${\bf v}(\rb)= \hat{e}_x v(y)$ is by 
prescription a function of $y$ only, and points in direction $x$. Furthermore, due to the symmetry of the system, $\rho$ is invariant in the $xz$ plane 
\cite {note}, $\rho(\rb,t)=\rho(y,t)$. Hence, one has 
\begin{equation}
\nabla\cdot \rho(y,t) \hat{e}_x v(y) =0,
\end{equation}
and Eq.~\eqref{ddft_current} is solved by the solution in absence of flow, i.e., for $t\to\infty$, it is solved by the equilibrium solution. Eq.~\eqref{ddft_current} can hence  not capture the changes in $\rho(y,t)$ caused by shear.
\subsection{The closure of Ref.~\cite{Brader_Kruger_2011}}
In Refs.~\cite{Brader_Kruger_2011,Brader_Kruger_2011_2} a semi-phenomenological theory has been proposed, in which the influence of the velocity field is accounted for by adding a
flow term into (\ref{ddft_current}), describing the ``rolling around'' of two particles in shear flow, i.e., the overdamped ``scattering process'' of two particles in flow. 
The term of Ref.~\cite{Brader_Kruger_2011} is based on dynamical considerations, minimizing hydrodynamic friction in the scattering event. In the following subsection we will introduce direct approximations to the 
aforementioned equations of statistical physics. Interestingly, one of our final results (see Eq.~\eqref{eq:HS} below), is identical to the corresponding result of Ref.~\cite{Brader_Kruger_2011}, except for an 
additional prefactor of $\pi/3\approx1.05$. As Eq.~\eqref{eq:HS} is the limiting result for large shear rates, and the description of the scattering event in Ref.~\cite{Brader_Kruger_2011} is purely deterministic, 
this agreement is reassuring, as one might expect that the statistical description becomes the more deterministic, the larger the shear rate is.
\subsection{Simple consistent closure}\label{sec:clo}
In this subsection, we collect conditions for a closure for $\rho^{(2)}(\rb,\rb')$ for sheared systems, and propose a simple form that obeys them.
Based on symmetry and force balance, we require: 
\begin{enumerate}
\item For vanishing shear rate $\dot\gamma$, the density distribution should approach the equilibrium one.\label{1}
\item Quantities that are antisymmetric in shear rate (e.g. the shear stress, compare Sec.~\ref{stress_tensor} for the stress tensor) are linear in $\dot\gamma$ for small rates, while symmetric quantities
(e.g. the diagonal components of the stress tensor) are at least of order $\dot\gamma^2$ for small rates. \label{2} 
\item In stationary situations, all volume elements are force free, such that interparticle forces,  Brownian forces and external forces are balanced \cite{dhont}. In the situation depicted in Fig.~\ref{fig:1}, 
the osmotic pressure exerted on the external potential (i.e. the wall) by the suspension must be equal to the corresponding stress tensor component (here $\sigma_{yy}$) in the suspension.\label{4}
\end{enumerate}
The structure in the bulk is characterized by the pair-correlation function,

\begin{equation}\label{g_definition}
 g(\rb-\rb')\equiv\frac{\rho^{(2)}(\rb,\rb')}{\rho(\rb)\rho(\rb')}\Bigl|_{bulk} \, .
\end{equation}
Previous computations for bulk systems \cite{dhont,Bergenholtz02,brady_morris} show that $g$ has the symmetries required by condition \ref{2}, hence using it in the
approximative closure is beneficial, as it automatically imprints these symmetries into the solution for the inhomogeneous system. A simple way to include it is via a {\it superposition approximation for the
distorted (shear-) part} of the pair correlation,
\begin{align}\label{superpos}
\rho^{(2)}_{neq}(\rb,\rb')\approx \rho(\rb)\rho(\rb')g_{neq}(\rb-\rb').
\end{align}
By definition, Eq.~\eqref{superpos} is valid in homogeneous, i.e., bulk systems, and is an approximation in inhomogeneous situations (that we aim for). We defined
$g_{neq} (\rb-\rb')\equiv g(\rb-\rb')-g_{eq}(|\rb-\rb'|)$, the nonequilibrium part of the pair correlation, where $g_{eq}(|\rb-\rb'|)$ refers to the system without shear. 
Noting that the total $\rho^{(2)}$ contains also the part that fullfills Eq.~\eqref{sum_rule} for the given (non-equilibrium) profile, we write the intuitive approximation,
\begin{multline}
\int\!d\rb'\, \rho^{(2)}(\rb,\rb')\nabla_{\rb} \phi(|\rb\!-\!\rb'|)\\
\approx \rho (\rb) \!\left(\!\!\nabla_\rb \frac{ \delta \mathcal{F}_{\rm ex}}{\delta \rho(\rb)}+\!\!\int \!\!\rho(\rb')  g_{neq}(\rb-\rb')\nabla_\rb \phi(|\rb-\rb'|) d \rb' \!\right). 
 \label{sum_rule_impr} 
\end{multline}
Again, $\mathcal{F}_{\rm ex}$ is the  excess free energy functional (compare Eq.~\eqref{sum_rule}). We have our final Shear-DDFT equation for the non-equilibrium current,
\begin{multline}
{\bf j}(\rb,t)\approx\rho(\rb,t)\vb(\rb)-\beta D \nabla V_{ext}(\rb)\rho(\rb,t)-\beta D \rho(\rb,t) \\
\times \left (\nabla\frac{\delta \mathcal{F}[\rho(\rb,t)]}{\delta \rho(\rb,t)} +\!\!\int \!\!\rho(\rb')  g_{neq}(\rb-\rb')\nabla_\rb\phi(|\rb-\rb'|) d \rb' \!\right).
\label{ddft_current_i}
\end{multline}
We note the important difference to equilibrium cases that Eq.~\eqref{ddft_current_i} contains {\it specific input from the dynamics}, namely the solution of the Smoluchowski equation for the pair-correlation
function $g_{neq}(\rb-\rb')$. By using this pair correlation function as an input, Eq.~\eqref{ddft_current_i} benefits from the adequate symmetry properties. The pair correlation function also provides the 
statistical physics description of a ``pair scattering'' event, thereby differing conceptually from the previous closure suggested in Ref.~\cite{Brader_Kruger_2011}. 

The distorted bulk pair-correlation function is an input into the theory. It can be computed by different means, e.g. using MCT \cite{Henrich07,Amann13}, or by use of Eq.~\eqref{ddft_current} together with a
tagged particle method \cite{Johannes_Joe_2013}. It can be computed analytically  in the limit of small particle density, in the two limits of small or large shear rates \cite{Bergenholtz02,brady_morris}. 
See Sec.~\ref{bp} for analytical estimates of $g$ used as input. 

The conditions \ref{1}-\ref{4} are indeed fulfilled by Eq.~\eqref{ddft_current_i}. Condition \ref{1} is fulfilled, since $g_{neq}$ vanishes at vanishing shear. Condition \ref{2} becomes evident in the explicit
discussion of hard spheres systems (Sec.~\ref{equations}), while we discuss condition \ref{4} in detail in Sec.~\ref{pressure}.

\section{Stress tensor}\label{stress_tensor}
We introduce the stress tensor as it is important for the analysis of rheological properties of the inhomogeneous system, as e.g. done in the stress ensemble in Sec.~\ref{ensembles} below.
\subsection{Exact relation}
The microscopic (colloidal) stress tensor ${  \boldsymbol \sigma}$ is defined in terms of the force $\bf F$ on a volume element due to particle interactions and Brownian forces,
\begin{align}
{\bf F}=-\nabla\cdot {  \boldsymbol \sigma} .
\end{align}
Noticing that Eq.~(\ref{ddft_current_0}) is a microscopic force balance, it can be rewritten  in terms of ${  \boldsymbol \sigma}$, to obtain the exact relation,
\begin{multline}
{\bf j(\rb,t)}=\rho(\rb,t)\vb(\rb)+\beta D \left(\nabla \cdot {  \boldsymbol \sigma}  - \nabla V_{ext} \right).
\label{force_balance}
\end{multline}

The force balance hence yields directly the divergence of $ {  \boldsymbol \sigma}$ by comparing Eq.~(\ref{force_balance}) with Eq.~(\ref{ddft_current_0}), 
\begin{multline}
\nabla \cdot {\bf {\boldsymbol \sigma}}(\vct{r})=-k_BT \nabla \rho(\rb)\\
 - \int d^3\rb' [\nabla_{\rb} \phi(|\rb-\rb'|)] \rho^{(2)}(\vct{r},\vct{r}').\label{divsigma}
\end{multline}

The tensor $ {\boldsymbol \sigma}$ itself is given by \cite{Kreuzer}, 
\begin{multline}
{\bf {\boldsymbol \sigma}}(\vct{r})=-k_BT\rho(\vct{r}) {\bf I}+\frac{1}{2}\int_0^1 d\lambda\int d^3 \vct{r'} \times \\
\times \frac{\vct{r'}\vct{r'}}{r'} [\partial_{r'} \phi(r')]\rho^{(2)}(\vct{r}+(1-\lambda)\vct{r'},\vct{r}-\lambda\vct{r'}) \:.\label{eq:sigmai}
\end{multline}
The divergence of Eq.~\eqref{eq:sigmai} agrees with Eqs.~\eqref{divsigma} and \eqref{ddft_current_0}, and Eq.~\eqref{eq:sigmai} approaches the more familiar form \cite{irving_kirkwood} for homogeneous systems,
\begin{align}
{\bf {\boldsymbol \sigma}}=-k_BT\rho_0 {\bf I}+\frac{1}{2}\rho_0^2\int d^3 \rb \frac{\vct{r}\vct{r}}{r} [\partial_r \phi(r)]g(\vct{r}) \:,\label{eq:sigmab}
\end{align} 
where $\rho_0$ is the density of particles in the bulk.

The first term on the right hand side of Eqs.~(\ref{divsigma}), (\ref{eq:sigmai}) or (\ref{eq:sigmab}) is entropic, i.e., related to thermal motion of the particles (the Brownian force \cite{dhont}), while the
second term is related to their mutual interactions.

Eq.~\eqref{divsigma} yields the (divergence of the) stress in terms of the density, and hence can be used to compute {\it rheological properties} in a consistent manner, after a specific closure for the 
two-particle density is employed. This result is hence general, and provides a strategy that is independent of the specific approximation proposed in Eq.~\eqref{sum_rule_impr}.  

\subsection{Approximate expression in Shear-DDFT}
By use of the approximate closure of Eq.~\eqref{sum_rule_impr}, we find the consistent expression for the stress by writing 
\begin{multline}
\nabla \cdot {\bf {\boldsymbol \sigma}}(\vct{r})\approx-k_BT \nabla \rho(\rb) - \rho(\rb) \Bigl(\nabla \frac{ \delta \mathcal{F}_{\rm ex}}{\delta \rho(\rb)} \\
+\!\!\int \!\!\rho(\rb')  g_{neq}(\rb-\rb')\nabla_{\rb}\phi(|\rb-\rb'|) d^3 \rb' \Bigl) \,.\label{divsigma2}
\end{multline}

\section{Contact value and osmotic pressure}\label{pressure}
In this section, considering the situation depicted in Fig.~\ref{fig:1},   we analyze the component of the pressure perpendicular to the wall, thereby discussing in more detail the condition 
\ref{4} of Sec.~\ref{sec:clo}.
\subsection{General properties} 
Recalling Eq.~\eqref{force_balance} for the particle current in terms of the particle stress,
\begin{multline}
{\bf j(\rb,t)}=\rho(\rb,t)\vb(\rb)+\beta D \left(\nabla \cdot {  \boldsymbol \sigma}  - \nabla V_{ext} \right),
\end{multline} 
and assuming a steady state in the coordinates defined in Fig.~\ref{fig:1}, we have 
\begin{equation}
j_y(\rb)=0 \, .
\label{ddft_impr} 
\end{equation}
As all the quantities in Eq.~\eqref{force_balance} are invariant in $x$ and $z$ directions \cite{dhont,Larson}, Eq.~\eqref{force_balance} simplifies with Eq.~\eqref{ddft_impr} to

\begin{equation}
-\int_{-\infty}^\infty dy \rho(y) \partial_yV_{ext}(y)=-\int_{-\infty}^\infty dy\partial_y \sigma_{yy}({y}) .
\label{pressure_derivation}
\end{equation}
The term on the l.h.s. is identified as the force per surface area exerted on the external potential by the particles, $F_y^{ext}/A=-\int_{-\infty}^\infty dy \rho(y) \partial_yV_{ext}(y)$. In case  $V_{ext}$ grows
sufficiently fast towards $y\to-\infty$ (e.g. in the case of an inpenetrable wall), we have vanishing density at $y=-\infty$, and hence 
\begin{equation}
\frac{F_y^{ext}}{A}=-\sigma_{yy}(\infty) .
\label{pressure_derivation_2}
\end{equation}
Eq.~\eqref{pressure_derivation_2} states the expected (and trivial) fact that the $yy$ component of the stress tensor in the bulk equals the force acting on the wall by the suspension, {\it Actio et Reactio}. 
For us it is important to understand if the approximation of Eq.~\eqref{sum_rule_impr} fulfills this principle. It does so due to the following; The relation for the stress in Eq.~\eqref{eq:sigmai} is
mathematically in agreement with Eqs.~\eqref{divsigma} and \eqref{ddft_current_0} for any $\rho^{(2)}(\rb,\rb')$ that is symmetric in the two arguments. Indeed, this symmetry is demanded by the 
physical definition of $\rho^{(2)}$, and we see here the importance of conserving  this symmetry when introducing approximations -- Eq.~\eqref{superpos}  does conserve it. Since, additionally, the approximative 
closure in Eq.~\eqref{sum_rule_impr} yields by construction the correct stress tensor components in the bulk (e.g. at $y\to\infty$), we have shown with Eq.~\eqref{pressure_derivation_2} that it reproduces the 
correct pressure of the suspension at the wall, i.e.,  the correct absorption at a wall, thus establishing consistency of pressures, as demanded in condition \ref{4} of Sec.~\ref{sec:clo}. 
Compare Fig.~\ref{fig:con} in Sec.~\ref{equations} below, where this is demonstrated explicitly.

\subsection{Specific properties for hard potentials -- contact value}
For the case of hard spheres of radius $R$ confined by a hard wall (located in the $y=0$ plane), additional properties can be found. Noting that in that case $\rho^{(2)}(\rb,\rb')$ vanishes if $y<R$ 
or $y'<R$, it is apparent that the second term in Eq.~\eqref{eq:sigmai} approaches zero for $y$ approaching $R$, and we have in that case,
\begin{equation}
{\bf {\boldsymbol \sigma}}(y=R)=-k_BT\rho(y=R) {\bf I}.\label{eq:sigmaii}
\end{equation} 
Noting furthermore that (directly from Eqs.~\eqref{ddft_impr} and \eqref{force_balance}), $\partial_y \sigma_{yy}=0$ for $y\geq R$ (where $V_{ext}=0$), we can extend Eq.~\eqref{eq:sigmaii} to the range of $y\geq R$,
\begin{equation}
{ \sigma}_{yy}(y\geq R)=-k_BT\rho(y=R) .\label{eq:sigmaiii}
\end{equation}
In particular, Eq.~\eqref{eq:sigmaiii} holds for $y\to\infty$, and we have shown, together with Eq.~\eqref{pressure_derivation_2}, that the pressure at the wall is indeed given by the density contact value,
in equilibrium and under the considered driving,
\begin{equation}
\frac{F_y^{ext}}{A}\equiv P_{particles}^{osm}=k_BT\rho(y=R).
\label{pressure_derivation_3}
\end{equation}
Eq.~\eqref{eq:sigmaii} also directly shows that the shear stress due to interactions vanishes at the wall,
\begin{equation}
 { \sigma}_{xy}(y)\Bigl|_{y=R}=0 \, .
\label{outermost_stress} 
\end{equation}
\section{Stress ensemble versus shear ensemble}\label{ensembles}
Using the approximate closure of Shear-DDFT in Eq.~(\ref{sum_rule_impr}) allows computation of the density profile $\rho(\rb)$, if the shear rate $\dot\gamma$ is given. Often, one may assume that the shear rate is
uniform throughout the system (which would be the case in the absence of Brownian particles). Such system (or theoretical computation) where the flow velocity or shear rate is prescribed may be termed
{\it shear ensemble}, in the notion of the well known concepts of equilibrium statistical physics; The microcanonical ensemble prescribes the internal energy, while in the canonical ensemble, temperature is the 
external control parameter. Most theoretical works on driven suspensions are in the framework of the shear ensemble, as e.g. the one of Ref.~\cite{Brader_Kruger_2011}, as well as the aforementioned references on 
MCT \cite{Fuchs02,pnas}.  

Especially in {\it inhomogeneous systems}, the physical (experimental) situation may be better described by a different ensemble, which we term the {\it stress ensemble}; It is characterized by the
circumstance that all volume elements are force free, meaning that the stress tensor (including all forces) has no divergence. We conclude that this includes the requirement that the shear stress $\sigma_{xy}$ 
must be constant throughout the system. For the setup of Fig.~\ref{fig:1}, we have to regard its $y$-derivative, which from Eq.~(\ref{divsigma}) is given in exact form, 
\begin{equation}
\frac {\partial}{\partial y}\sigma_{xy}(y)=-\int d^3\rb' \frac{x'}{r'} [\partial_{r'} \phi(r')] \rho^{(2)}(\vct{r},\vct{r}+\vct{r}') \, .\label{eq:divsigmaSTRESS}
\end{equation}
From the above argument, we hence require that this derivative must vanish. There is however yet another contribution to the shear stress, due to the (bare) solvent viscosity, which  can be written as
$\sigma_{xy}^{(s)}=\dot\gamma(y)\, \nu$,
where $\nu$ is the coefficient of viscosity. In the stationary situation, we hence demand the total shear stress to be uniform, and with Eq.~(\ref{eq:divsigmaSTRESS}), 

\begin{align}
0=&\frac {\partial}{\partial y}\sigma_{xy}(y)+ \frac {\partial}{\partial y}\sigma^{(s)}_{xy}(y)\notag\\
=&-\int d^3\rb' \frac{x'}{r'} [\partial_{r'} \phi(r')] \rho^{(2)}(\vct{r},\vct{r}+\vct{r'}) \;+\;\nu \frac {\partial}{\partial y} \dot\gamma(y) .
\label{velocity_eq_general}
\end{align}
Compare also the comment of Sec.~\ref{sec:hyd} on hydrodynamic interactions.
The coefficient of viscosity $\nu$ is related to  diffusion coefficient $D$ via the Einstein relation, e.g. for hard spheres of radius $R$,
\begin{equation}
\nu=\frac{k_BT}{6\pi D R} . \label{einstein}
\end{equation}
Equations (\ref{ddft_current_0}) and (\ref{velocity_eq_general}) determine the density profile in the stress ensemble. Note that the shear rate is not uniform in the stress ensemble, as it is an output of the
theory. Nevertheless, for the situation of Fig.~\ref{fig:1} it tends to the bulk value $\dot \gamma_0\equiv \dot\gamma(y\to\infty)$ far from the wall in the absence of layering transitions in the bulk, compare
Ref.~\cite{Brader_Kruger_2011}). 

The explicit (approximate) form of Eq.~\eqref{velocity_eq_general} for hard spheres is given in Eq.~\eqref{velocity_eq2} below.

\section{Specific forms and results for hard spheres}\label{equations}
In this section, we finally evaluate explicitly (including numerically) the exemplary situation depicted in Fig.~\ref{fig:1}, i.e., a hard sphere suspension confined by a hard wall, and sheared parallel to the wall. 
\subsection{Bulk pair correlation}\label{bp}
The pair correlation function is an input to Shear-DDFT (compare Eq.~\eqref{sum_rule_impr}), and we summarize in this subsection known analytical forms used in the following. It can be computed analytically for 
the limits of small or large shear rates, considering the limit of low particle density. 
In the limit of small shear rates, one has the typical quadrupolar form \cite{mk_diploma},
\begin{align}
\lim_{\rho_0\to0,\dot\gamma_0\to 0} g_{neq}(\rb)\Bigl|_{r=2R}=- \frac{1}{6} \dot\gamma_0 \frac{xy}{D}. \label{g_small_shear}
\end{align}
The derivation of Eq.~(\ref{g_small_shear}) from \cite{mk_diploma} is reproduced in Appendix B. In the limit of large rates, one has \cite{brady_morris}
\begin{align}
\lim_{\rho_0\to0,\dot\gamma_0\to\infty}g_{neq}(\rb)\Bigl|_{r=2R}=- \frac{1}{6}\dot\gamma_0 \frac{xy}{D} \Theta(- x y).\label{eq:g}
\end{align}
The unit step function $\Theta$ reflects the circumstance that the contact value of $g$ increases linearly with shear on the up-stream sides, but is of order unity on the downstream sides. For finite densities,
exact analytical results are not available, and approximative numerical methods can be used to find $g_{neq}$ \cite{Henrich07,Johannes_Joe_2013,Amann13}. In order to keep an analytical form for the Shear-DDFT 
equation, we use a simple extrapolation to higher densities (e.g. similarly suggested in Ref.~\cite{brady_morris}) by use of the contact value in equilibrium, 
\begin{align}
\lim_{\dot\gamma_0\to 0} g_{neq}(\rb)\Bigl|_{r=2R}&\approx- \frac{1}{6} \dot\gamma_0\, \frac{xy}{D} g_{\rm eq}(2R), \label{eq:gs}\\
\lim_{\dot\gamma_0\to\infty}g_{neq}(\rb)\Bigl|_{r=2R}&\approx- \frac{1}{6}\dot\gamma_0\, \frac{xy}{D} \Theta(- x y) g_{\rm eq}(2R).\label{eq:gl}
\end{align}
The equilibrium contact value is in turn found from the very accurate Carnahan-Starling expression \cite{HansenMcDonald} ($\Phi=4/3 \, \pi R^3 \rho_0$ is the packing fraction), 
\begin{align}
g_{\rm eq}(2R)\approx\frac{1-\frac{\Phi}{2}}{(1-\Phi)^3}.
\end{align}
With this, the Shear-DDFT equation in Eq.~\eqref{ddft_current_i} can be solved numerically. In the following subsections, we use Eq.~\eqref{eq:gs} for the small shear rate, i.e for the linear response case,
and take Eq.~\eqref{eq:gl} for the finite rates shown (owing the small inconsistency that use of Eq.~\eqref{eq:gl} should be restricted to very large rates to the advantage of an analytical expression for the 
Shear-DDFT closure).
\subsection{Shear-DDFT equation}
The Shear-DDFT equation now follows directly from  Eqs.~(\ref{ddft_current_0}), \eqref{sum_rule_impr}, \eqref{eq:gs} and \eqref{eq:gl}. For small rates, using Eq.~\eqref{eq:gs}, we find that the shear rate
dependent term in Eq.~\eqref{sum_rule_impr} vanishes since the integral yields exactly zero due to symmetries. To first order in shear rate, the density distribution is thus
unchanged, as required by symmetry. It is thus evident that Eq.~\eqref{sum_rule_impr} incorporates this symmetry (compare condition \ref{2} in Sec.~\ref{sec:clo}). For large rates, $\dot\gamma\to\infty$, the 
Shear-DDFT approximation yields for hard spheres,
\begin{multline}
0=\beta \frac{\partial}{\partial  y}\left(\frac{\delta \mathcal{F}}{\delta \rho( y)} +V_{ext}( y)\right)+\frac{(2R)^2 Pe}{3}\, g_{eq}( 2R) \\ 
\times \int_{ y-2R}^{ y+2R} d y' \rho( y')\,  S(y,y') {\rm sgn}( \Delta)\, \Delta^2 \sqrt{1-\Delta^2}.
\label{eq:HS}
\end{multline}
with $\Delta\equiv (y'-y)/2R$, and the Peclet number $Pe\equiv\dot\gamma_0 (2R)^2/D$. $Pe$ characterizes the activity of Brownian motion as compared to the external driving.
 We included the possibility for a spatially varying rate $\dot\gamma(y)$ as a prerequisite for the stress ensemble below (one may also imagine situations
where the rate is a priori spatially dependent, e.g. in a Poiseuille flow). For a spatially varying rate, an additional step is necessary after introduction of Eq.~\eqref{sum_rule_impr}, (as the local rate can
vary over a particle diameter), and we introduce the function $S(y,y')$ in Eq.~\eqref{eq:HS} which is defined by,
\begin{equation}
S(y,y')=\frac{1}{y'-y} \int\limits_{y'}^y ds \frac {\dot\gamma(s)}{\dot\gamma_0}. 
 \label{shear_of_v} 
\end{equation}
$S(y,y')$ is constructed such that two interacting particles have the proper velocity difference according to their spatial displacement along $y$. For homogeneous shear rate, we obviously have
$S(y,y')=1$. In this case, Eq.~(\ref{eq:HS}) is equal to the corresponding equation in Refs.~\cite{Brader_Kruger_2011} and \cite{Brader_Kruger_2011_2}, however its second term is multiplied by $\pi/3$.
Up to this different prefactor, the result of Ref.~\cite{Brader_Kruger_2011} coincides with the one obtained now, however only in the limit of large shear rates.

\begin{figure}
\includegraphics[width=1\linewidth]{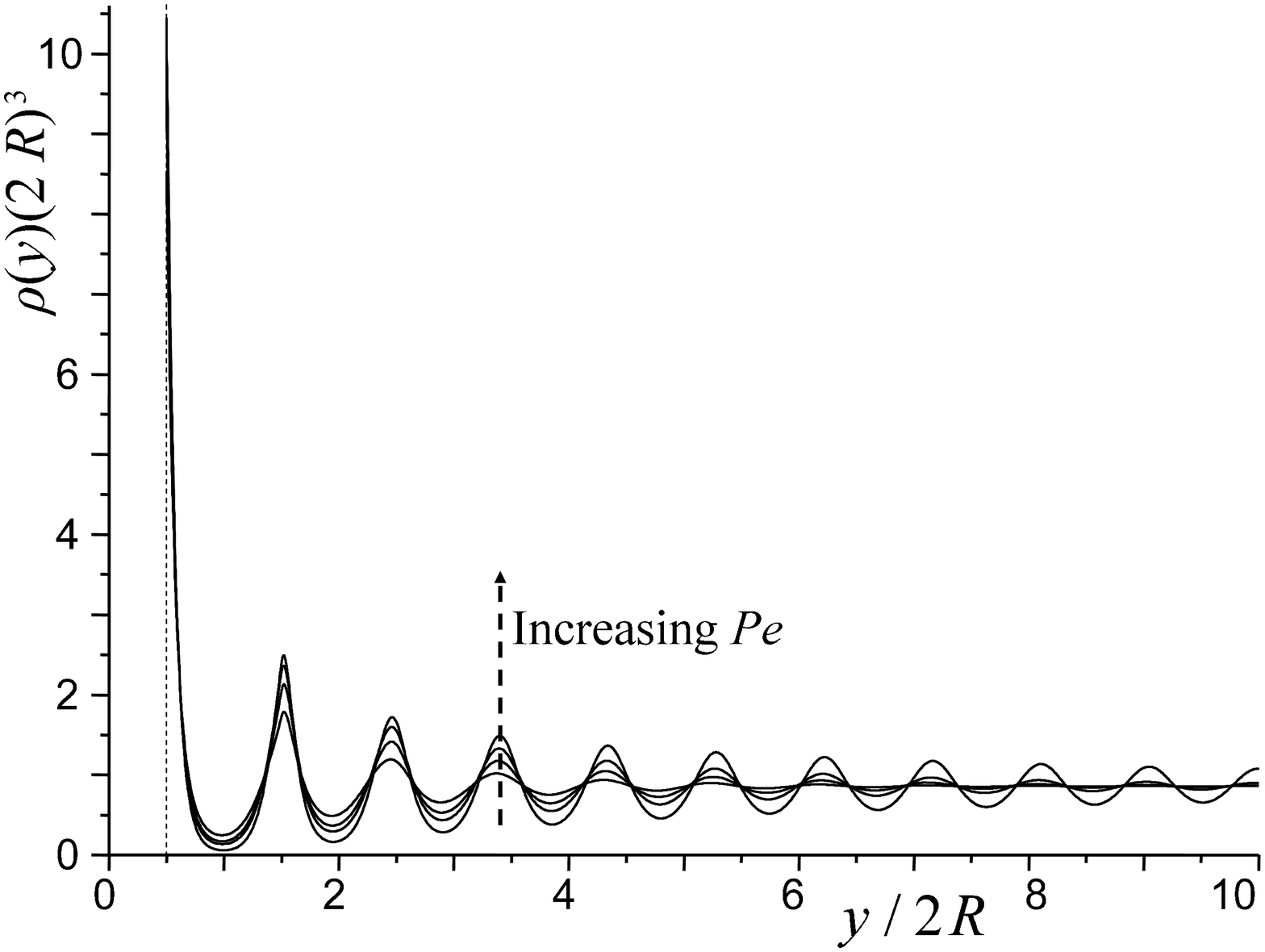}
\caption{\label{fig:rho}Density profiles for hard spheres opposite a hard wall with shear, obtained in the shear ensemble, i.e., from Eq.~\eqref{eq:HS} assuming a homogeneous shear rate. $Pe=0$, $6.26$, $9.39$,
$12.53$ and $\Phi=0.45$. The curve for smallest shear rate represents both equilibrium and linear response. The curves for finite shear rates are (up to the mentioned prefactor of $\pi/3$ in shear rate) identical
to the ones obtained in Ref.~\cite{Brader_Kruger_2011}.}
\end{figure}
Figure \ref{fig:rho} shows the resulting density profiles obtained from Eq.~\eqref{eq:HS} by use of a homogeneous shear rate for Peclet numbers ranging from $0$ to $12.53$.
The equilibrium curve and the linear response curve are identical.

\begin{figure}
\includegraphics[width=1\linewidth]{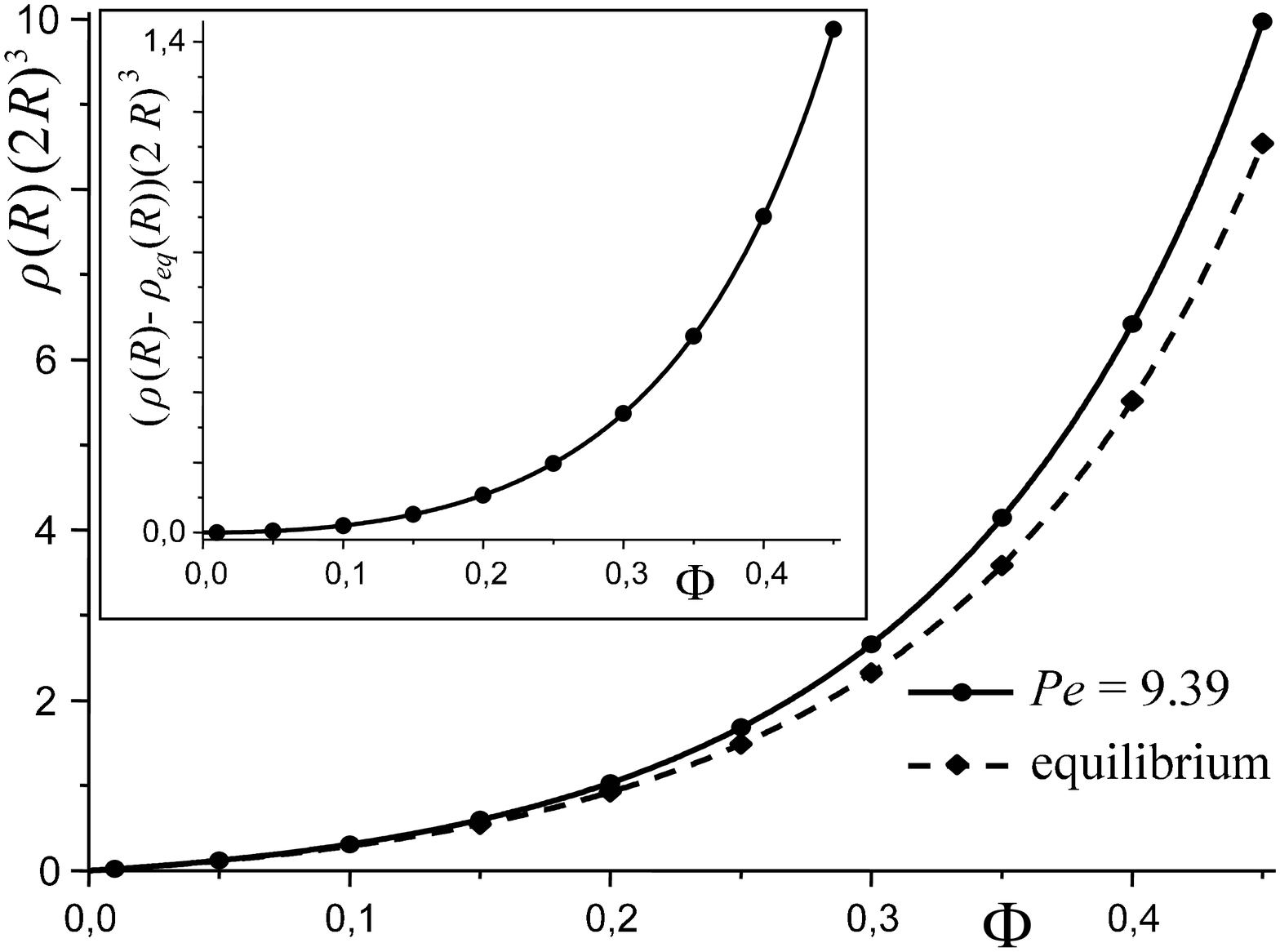}
\caption{\label{fig:con}Contact value of the density at the wall as a function of bulk packing fraction, for the equilibrium case, and for $Pe=9.39$.  
Lines are the analytical expressions of Eqs.~\eqref{equilibrium_contact_value_0} and \eqref{equilibrium_contact_value} (obtained from the bulk pressure), while data points are numerically computed contact values. 
Inset shows that the effect of shear on the contact value is of order $\Phi^2$ for small $\Phi$, as it is an effect due to particle interactions and thus absent for an ideal gas.}
\end{figure}
Figure \ref{fig:con} shows the contact value, i.e., $\rho(y=R)$ in Fig.~\ref{fig:rho}, as a function of bulk packing fraction. By use of the Rosenfeld functional (see Appendix A), one obtains the familiar result
in equilibrium, which is identical to the one of linear response, 
\begin{equation}
\lim_{\dot\gamma_0\to0}\rho(y=R)=\rho_{0}\frac{\left(1+\Phi+\Phi^2\right)}{\left(1-\Phi\right)^3} +\mathcal{O}(\dot\gamma^2).\label{equilibrium_contact_value_0}
\end{equation}
By use of the closure for large rates in addition, i.e., Eq.\eqref{eq:gl}, we obtain the contact value for large shear rate,
\begin{equation}
\lim_{\dot\gamma_0\to\infty}\rho(y=R)=\rho_{0}\frac{\left(1+\Phi+\Phi^2\right)}{\left(1-\Phi\right)^3} +\rho_{0} \frac{4 Pe }{15\pi}\frac{1-\frac{\Phi}{2}}{(1-\Phi)^3}\Phi.\label{equilibrium_contact_value}
\end{equation}
Eqs.~\eqref{equilibrium_contact_value_0} and \eqref{equilibrium_contact_value} have been computed from Eq.~\eqref{eq:sigmaiii}, i.e., they give the stress tensor element $\sigma_{yy}$ in the bulk. In the figure,
we show that these bulk values indeed agree with the numerically obtained contact values (i.e., the contact values in Fig.~\ref{fig:rho}), thus demonstrating that Shear-DDFT yields a consistent contact value, as
demanded by item \ref{4} in Sec.~\ref{sec:clo}.

We finally note that Eqs.~\eqref{equilibrium_contact_value_0} and \eqref{equilibrium_contact_value} are, due to this consistency of pressures,  {\it independent}  of the ensemble, i.e., they also hold in the shear
ensemble, described in more detail below. For all other values of $y$, the density profiles in shear and stress ensemble differ in principle, this (not shown) difference being however almost invisible on the scale 
of Fig.~\ref{fig:rho}.

\subsection{Shear stress}
By substituting Eq.~\eqref{eq:gs}, or Eq.~\eqref{eq:gl}, into the general expression for stress, Eq.~\eqref{divsigma2},  we obtain consistently the local particle shear stress (its gradient),
\begin{multline}
\frac {\partial}{\partial y}\sigma_{xy}(y) 	\approx \frac{\alpha R }{3} \pi\,  k_BT \, Pe \, g_{eq}(2R) \rho(y) \\
\times\int_{y-2R}^{y+2R} d y' \left(1- \Delta^2\right) \rho( y') S(y,y')\Delta .
\label{velocity_eq}
\end{multline}
Here we included the prefactor $\alpha$, which takes the value $\alpha=1$ for low shear rates, i.e., if Eq.~\eqref{eq:gs} is substituted, and $\alpha=\frac{1}{2}$ for large rates, i.e.,  if Eq.~\eqref{eq:gl} is
substituted.  We see that the local shear stress is indeed linear in $\dot\gamma$
for small $\dot\gamma$, as required by condition \ref{2} in Sec.~\ref{sec:clo}. 

\begin{figure}
\includegraphics[width=1\linewidth]{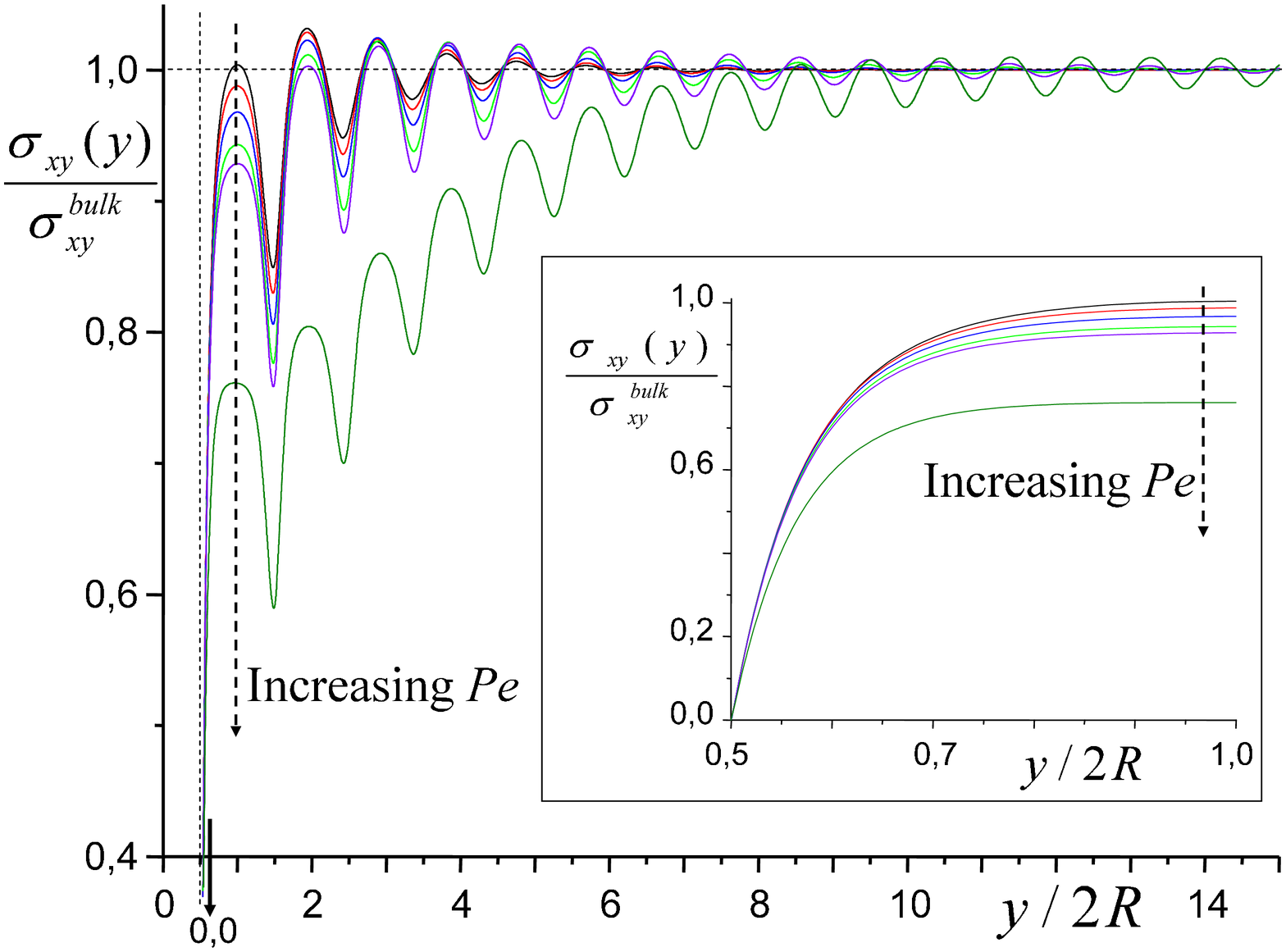}
\caption{\label{fig:sig}Shear stress $\sigma_{xy}$ as a function of distance to the wall, evaluated by use of Eq.~\eqref{velocity_eq} and the density profiles of Fig.~\ref{fig:rho}. Inset shows an amplification of 
the region close to the wall where the particle stress vanishes continuously. The different curves correspond to $Pe=0$, $3.13$, $6.26$, $9.39$, $10.95$, $12.53$ and $\Phi=0.45$.
}
\end{figure}

Fig.~\ref{fig:sig} shows the stress, resulting from Eq.~\eqref{velocity_eq} by use of the density profiles of Fig.~\ref{fig:rho}. It approaches the bulk value for large distances from the wall, and deviates from 
the bulk value in the proximity of the wall. Near the wall, the stress shows oscillations on the scale of the particle size, being small on the density peaks in Fig.~\ref{fig:rho} and large in the density valleys: 
Particles on the peaks slide by each other more easily than those in the valley. Additionally, there is an overall trend towards smaller stress on approach of the wall, showing that the layered structure near the
 wall has a smaller overall shear resistance. The decrease of the (relative) stress towards the wall increases with shear rate, since shear increases the layering structure. In the inset, we focus on the region 
close to the wall, where we explicitly show that the numerical results, in agreement with Eq.~\eqref{outermost_stress}, vanish for $y\to R$.  
Noticing that a varying stress component $\sigma_{xy}$ is unphysical in a steady state, we evaluate in the next subsection the stress ensemble, where the total stress is required to be homogeneous.

\subsection{Determination of the velocity profile in the stress ensemble}

By use of Eq.~(\ref{einstein}), Eq.~(\ref{velocity_eq_general}) for stress homogeneity reads for hard spheres 
\begin{multline}
\frac{\alpha(2R)^4}{2} \pi^2 \, g_{eq}(2R) \rho(y)\int_{y-2R}^{y+2R} d y' \left(1- \Delta^2\right) \\
\times \rho( y')S(y,y')\Delta +\frac{\partial  \dot\gamma( y) }{\partial { y} }\frac{1}{\dot\gamma_0} =0 .
\label{velocity_eq2}
\end{multline}  
Equations \eqref{eq:HS} and (\ref{velocity_eq2}) are used to determine the shear rate $\dot\gamma(y)$ and the density $\rho(y)$ by numerical iteration. Since Eq.~\eqref{velocity_eq2} contains the derivative of the
shear rate, knowledge of $\dot\gamma(y)$ at one point is required, and, for the semi-infinite system, we fix it far away from the wall to the value of the bulk, i.e., we require
$\lim_{y\to\infty}\dot\gamma(y)=\dot\gamma_0$. This yields the solution for $y\geq R$. For $0\leq y < R$, the rate is  constant, and fixed by 
requiring that the stress in this range is equal to the stress for $y\geq R$. As it is required to be the same everywhere, it can be computed from the bulk expressions, i.e., by Eqs.~\eqref{eq:gs} and 
\eqref{eq:gl}, together with Eq.~\eqref{eq:sigmab}  \cite{brady_morris},
\begin{equation}
\sigma_{xy}^{bulk}=\alpha \frac{8 \pi}{45}Pe\,\rho_0^2\,R^3\,k_BT\,g_{eq}(2R) ,
 \label{sigma_bulk_low_shear} 
\end{equation}
From this equation and Eq.~(\ref{einstein}), we can determine the shear rate $\dot \gamma_g$ in the gap, i.e., for $0\leq y < R$,
\begin{equation}
\frac{\dot \gamma_g-\dot\gamma_0}{\dot\gamma_0}=\frac{12\alpha}{5}\,\Phi^2   g_{eq}(2R)  .
 \label{shear_gap} 
\end{equation}
This procedure determines the shear rate for $y\geq0$ uniquely, and as we have  $\dot\gamma(y)=\partial_y v(y)$, the velocity $v(y)$ is determined up to a constant. This constant can be fixed by assuming that 
$v(y)$ vanishes at the wall, i.e., assuming no-slip boundary conditions for the solvent. 

\begin{figure}
\includegraphics[width=1\linewidth]{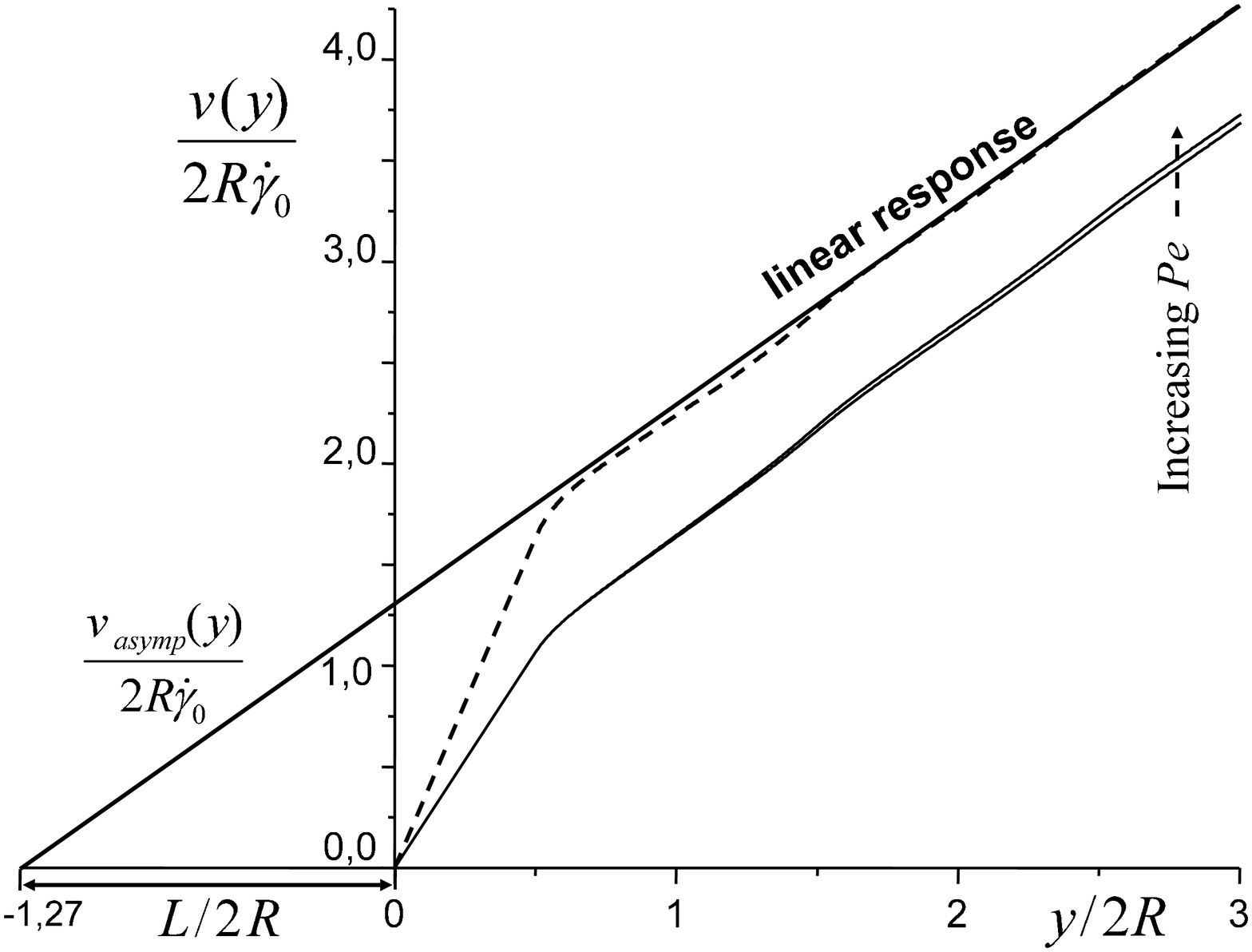}
\caption{\label{fig:vel}Solvent velocity $v$ as a function of distance from the wall, for different shear rates, obtained in the framework of the stress ensemble. The strongest deviation of the velocity from the
naively expected straight line is given in the gap $0\leq y < R$, and oscillations are visible in the region where particles form layers. Dashed curve corresponds to the linear response case. The adjacent to it 
straight line is the asymptote for $y\to\infty$, and used to construct the slip length $L$. The two neighboring solid curves correspond to $Pe=6.26$, and $Pe=10.45$. For all curves, $\Phi=0.45$.    
}
\end{figure}
Fig.~\ref{fig:vel} shows the resulting velocity for the parameters used in the previous figures. We see that the velocity deviates from the straight line, its slope showing a strong deviation in the gap 
$0\leq y < R$, where no particle centers are present, and the stress is purely due to the solvent. For $y\geq R$, the velocity shows oscillations which can be related to the oscillations of the stress in 
Fig.~\ref{fig:sig}. Fig.~\ref{fig:vel} also shows the construction of the slip length $L$ by inclusion of the asymptote that is approached for $y\to\infty$ (compare also Fig.~\ref{fig:1}). 

\begin{figure}
\includegraphics[width=1\linewidth]{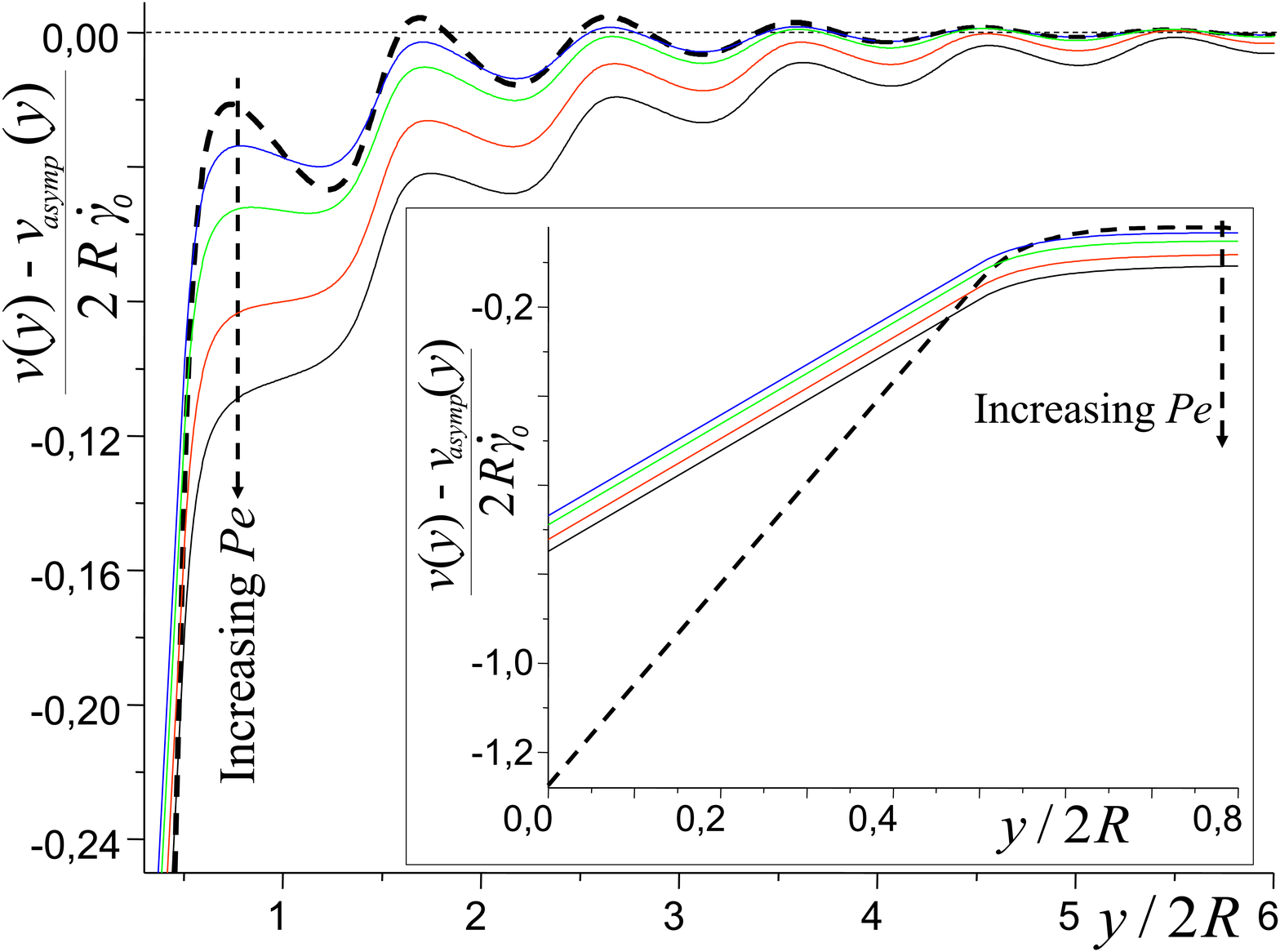}
\caption{\label{fig:veld}Same data as in Fig.~\ref{fig:vel}, after subtraction of the asymptotic velocity. We see the oscillatory nature of the velocity. Interestingly, not the velocity, but its gradient is in 
phase with the density oscillations shown in Fig.~\ref{fig:rho}. Inset shows the full range on the vertical axis. Dashed line corresponds to the linear response. Solid lines correspond to $Pe=3.13$, $6.26$, 
$9.39$, $10.95$. For all curves, $\Phi=0.45$.
}
\end{figure}

\begin{figure}
\includegraphics[width=1\linewidth]{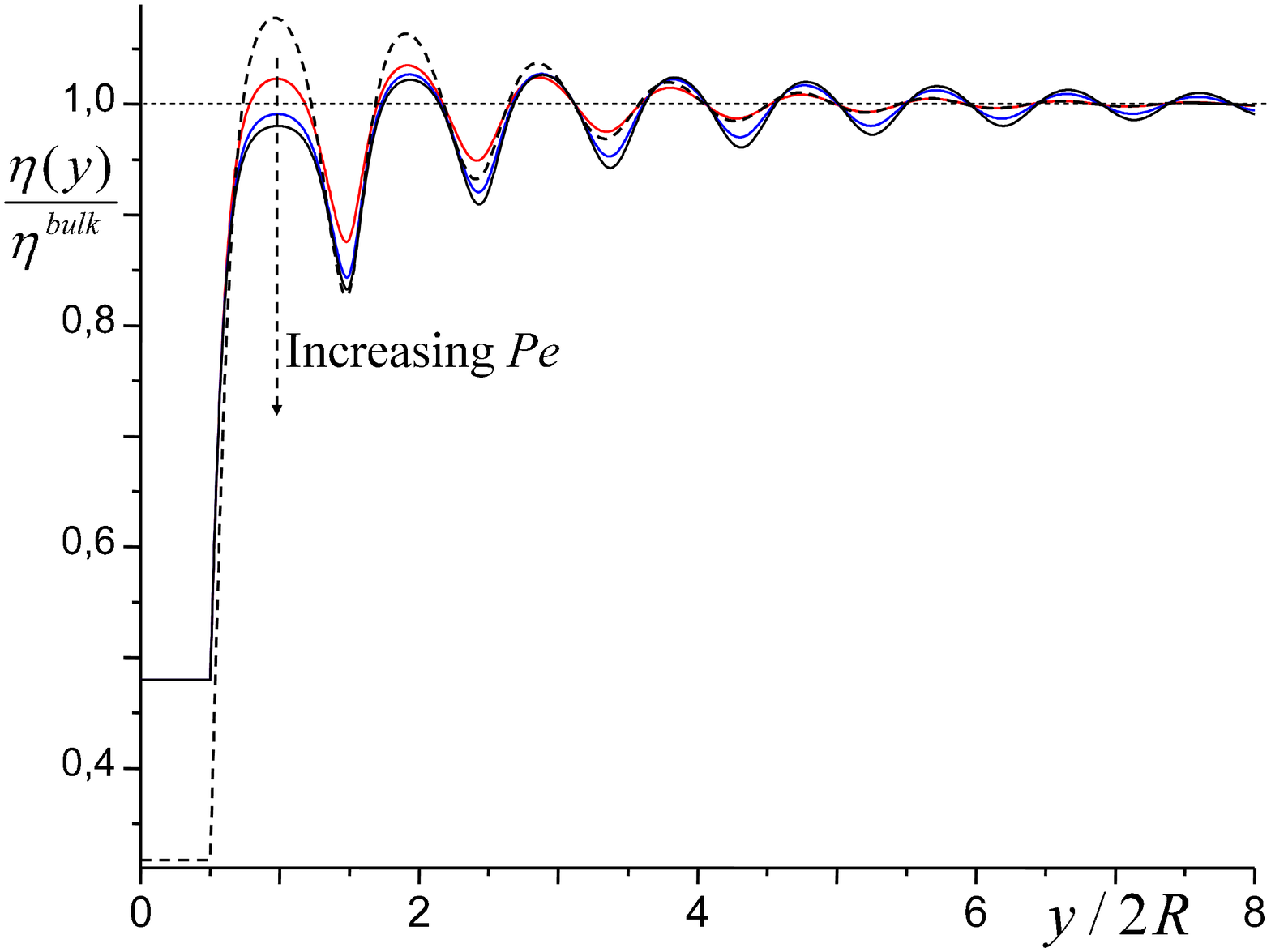}
\caption{\label{fig:vis}Inverse of the local shear rate $\eta(y)\equiv \dot\gamma^{-1}(y)$, which may be interpreted as the local shear viscosity. It is apparent that this viscosity is maximal in the valleys of the 
density profiles in Fig.~\ref{fig:rho}, and minimal on the peaks. This can be interpreted by the easier sliding of layers at the peaks. Dashed line corresponds to linear response. Solid lines correspond to
$Pe=3.13$, $9.39$, $10.95$. For all curves, $\Phi=0.45$.
}
\end{figure}

Fig.~\ref{fig:veld} focuses on the oscillatory part of the velocity, by displaying the deviation from the asymptote shown for the linear response case in Fig.~\ref{fig:vel}. We note that the oscillatory behavior 
of the velocity is  out of phase with respect to the  density oscillations. As illustrated in more detail below (see Fig.~\ref{fig:vis}), it is the {\it gradient} of the velocity (i.e. the shear rate) that is in 
phase with the density; It is high in the peaks and low in the valleys. In Fig.~\ref{fig:veld}, both $v$ and its gradient, the shear rate, are continuous at $y=R$. This continuity is worth mentioning, as it
is not a numerical coincidence. It follows directly from the continuity of the particle shear stress at $y = R$, see Eq.~\eqref{outermost_stress}. 

In Fig.~\ref{fig:vis} we show the inverse of the shear rate $\dot\gamma^{-1}$, normalized to the bulk value. It may be {\it interpreted} as the local suspension viscosity 
$\eta(y)$. We see that this quantity is indeed in phase with the density, being low on the peaks (as the particles at the peaks are more mobile) and high in the valleys. Future work might establish possible
 relations to other measurable quantities, as e.g. the (local) diffusivity of particles \cite{Lang10}, which is readily accessible in experiments \cite{Mittal08} and shows oscillatory behavior as a function of the
distance to the wall as well. 

\begin{figure}
\includegraphics[width=1\linewidth]{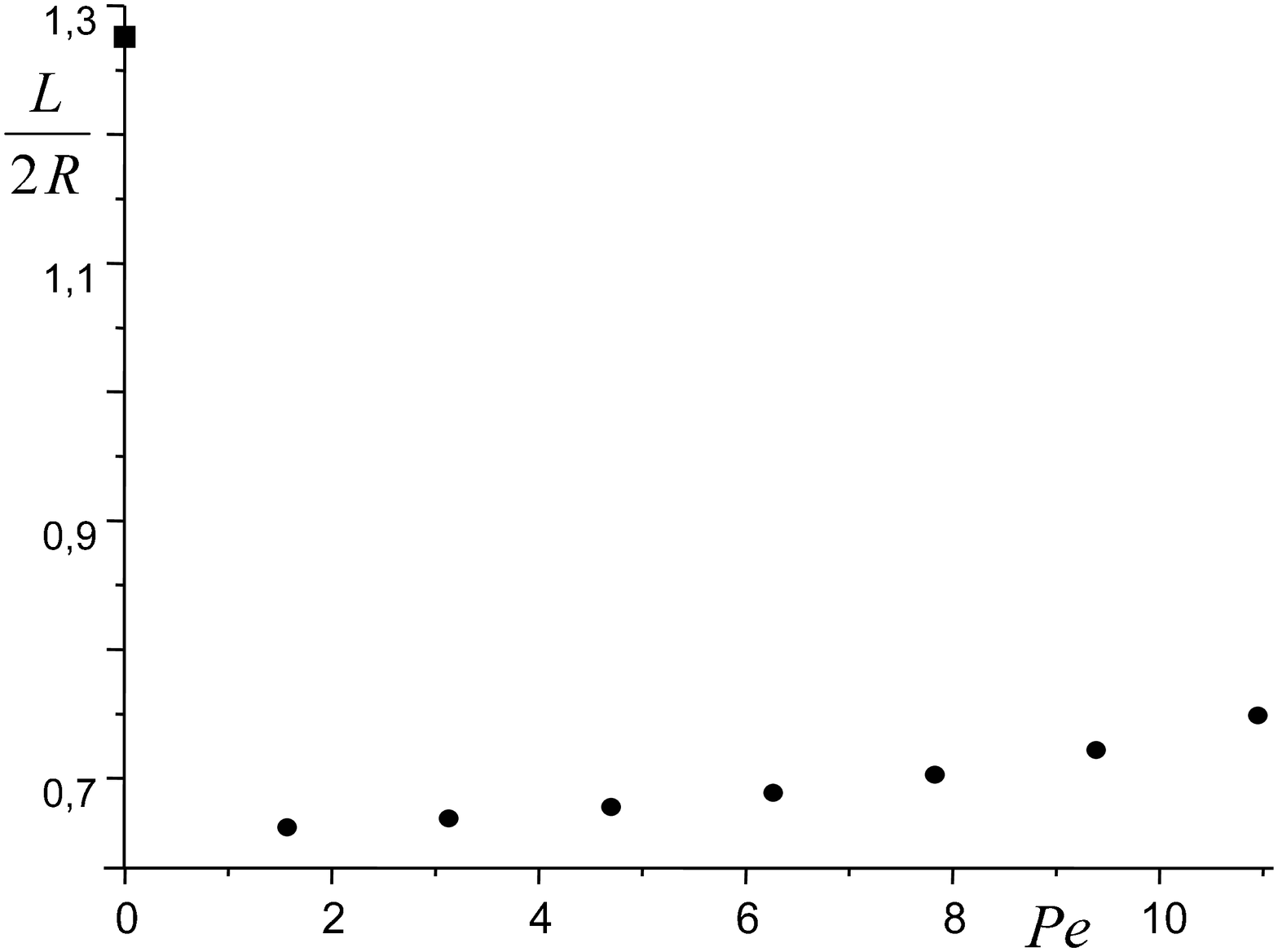}
\caption{\label{fig:slip}Slip length $L$ as a function of shear rate. The square shows the linear response result, while circles show the result obtained with the closure valid for large $\dot\gamma_0$. 
The packing fraction is $\Phi=0.45$.
}
\end{figure}

Fig.~\ref{fig:slip} finally displays the slip length $L$, an example of its construction is shown in Fig.~\ref{fig:vel}. This length is of interest, as it has direct consequences on the behavior of  complex fluids
in confined systems (nanofluidics), e.g. the fluid flow in a narrow channel will be faster than expected from the suspensions' bulk viscosity. When considering flow in small geometries, as a rough estimate, one 
may treat the complex fluid as a continuum with corresponding slip boundary conditions. 

The slip length is large in the linear response regime (small $Pe$), decreases then as a function of bulk shear rate, and finally slowly increases towards larger rates. The details of the behavior for intermediate 
rates ($Pe\approx 1$) cannot be studied by use of the analytical estimates for the pair correlation in Eqs.~\eqref{eq:gs} and Eqs.~\eqref{eq:gl}, and we leave this for future work (e.g. by use of numerical 
results for $g_{neq}(\rb)$ as input).

\section{Summary}\label{results}
We investigated a {\it confined colloidal suspension under shear} by means of density functional theory. Apart from a specific closure approximation for shearing, we investigated general symmetry properties as 
well as the behavior of the different elements of the stress tensor. Noting that the resulting off-diagonal element corresponding to the shear stress is not necessarily homogeneous at a homogeneous shear rate, we
introduced a second equation, additional to the one determining the density profile. The second equation requires homogeneity of shear stress and is used to determine in addition the profile of solvent
velocities, as this cannot be assumed to be known in confinement.

We analyzed the exact expression for the {\it stress tensor} for the non-equilibrium situation depicted in Fig.~\ref{fig:1} and showed that the suggested approximative closure generally obeys Newton's Actio et 
Reactio, i.e., the force acting on the external potential equals the corresponding component of the bulk pressure. Specifically, for the case of hard potentials, we show that the density contact value at the wall
indeed equals the particle pressure acting on the wall, a result which to our knowledge had been restricted to equilibrium. Furthermore, the shear stress produced by the particles vanishes at approaching contact
with the wall.
 
The oscillations of particle density near the wall become more pronounced due to shear, and lead to {\it inhomogeneous response to shear}, as seen in the resulting solvent velocity near the wall: It oscillates as 
well. In particular, the inverse of local shear rate (interpreted as a local viscosity) is low on the peaks of density and high in the valleys. 

Specifically, the analytical expressions for Shear-DDFT are valid for high shear rate, and for linear response (vanishing shear). The resulting slip length is locally maximal at linear response, then drops for
increasing rates, and shows a slow increase in the region of large shear rates. It is generally around one particle diameter.
\section{Acknowledgments}
We thank J.~M.~Brader, R.~Roth, R.~Evans, T.~Voigtmann, M.~Fuchs, and M.~Schmidt for useful discussions. This research was supported by Deutsche Forschungsgemeinschaft (DFG) grant No. KR 3844/2-1.
A.A. thanks also European Union International Research Staff Exchange Scheme (EU IRSES) grant No. PIRSES-GA-2010-269139. 

\section*{Appendix A. The Rosenfeld functional}\label{App:Ro}

The density $f_{ex}^{hs}$ of the Rosenfeld excess free energy $\mathcal{F}_{\rm ex}=\int \!d\rb f_{ex}^{hs}(\rb) $ for monodisperse hard spheres is expressed as
\begin{eqnarray}
\beta f_{ex}^{hs} [\rho] &=& 
-n_0\ln(1-n_3)  
\,+\,\frac{4\pi R\left(n_0^2 \, 4\pi R^2-|\boldsymbol{n}_1|^2\right)}{1-n_3} \notag\\
&&+\,\frac{8 \pi^2}{3}\frac{R^4 \, n_0 \left(n_0^2 R^2 - 3 |\boldsymbol{n}_1|^2\right)}
{(1-n_3)^2} \, ,
\label{rosenfeldfunctional}
\end{eqnarray}
where the three weighted densities are given by convolutions of the density profile with the corresponding weight functions, 
\begin{equation}
n_{\alpha}(\mathbf{r})=\int d\mathbf{r'}\rho(\mathbf{r'})
\,\omega^{(\alpha)}(\mathbf{r}-\mathbf{r'}) \, ,
\label{weighted_density}
\end{equation}
\begin{eqnarray}
\omega^{(3)}(\mathbf{r})&=&\Theta(R-r),\notag\\
\omega^{(0)}(\mathbf{r})&=&\frac{\delta(r-R)}{4\pi R^{2}},\notag\\
\boldsymbol{\omega}^{(1)}(\mathbf{r})&=&\frac{\mathbf{r}}{r}\frac{\delta(r-R)}
{4\pi R} \, .
\label{weightfunctions}
\end{eqnarray}

\section*{Appendix B. Bulk pair-distribution function of a shear flow of hard spherical particles in the limit of small shear rate and small particle density}
Let us consider the frame comoving with one of the particles of a homogeneous colloidal suspension having particles density $\rho_0$, the particle center is at $\rb=0$. 
The particle plays the role of a fixed obstacle for the other ones involved in both the driven flow and the thermal motion.

To a first approximation, in the comoving frame, the shear flow is not distorted by the fixed particle:
\begin{equation}
v_x=v_z=0\, , \,\,\,\,\,\,\,\,\,\,\,v_y=\dot \gamma_0 x \, ;
 \label{flow} 
\end{equation}
the pair density is not distorted by the obstacle, i.e. $\rho^{(2)}(\rb,\rb')$ is point-symmetrical with respect to $\rb-\rb'$; and the thermal motion is a normal one characterized by
the diffusion coefficient $2D$, where $D$ is the diffusion coefficient of the particles in the static frame. The last statement would be completely rigorous if there were only one particle except the fixed one. 
 
The particles density in the comoving frame can be expanded in a power series of the small shear rate, up to the first order: 
\begin{equation}
\rho(\rb)=\rho_0+\dot\gamma_0 \, \rho_1(\rb) \, .
 \label{density_expansion} 
\end{equation}

By substituting (\ref{density_expansion}) into (\ref{ddft_eq}), with taking into account the aforementioned approximations, one obtains that in the steady state $\Delta \rho_1(\rb)=0$. One has to choose
the solution of this Laplace equation, which fulfills the boundary condition, that requires the vanishing of the particles current, eq. (\ref{ddft_current_0}), through the surface of the obstacle:
\begin{equation}
\left( v_r(\rb)\rho_0-2D \, \dot\gamma_0 \frac{\partial}{\partial r}\rho_1(\rb)\right)\Bigl|_{r=2R}=0 \,.
 \label{boundary_condition} 
\end{equation}

$\rho(\rb)\rho_0$ in comoving frame is equal to $\rho^{(2)}(\rb',\rb'+\rb)=g(\rb)\rho_0^2$ in the static frame, where $\rb'$ is the position of the origin of comoving frame in the static frame. 
Thereby, one obtains the equation (\ref{g_small_shear}).


\begin{thebibliography}{43}
\bibitem{note} We assume the absence of transitions that could cause spontaneous symmetry braking in the $xz$ plane.
\end{thebibliography}

%

\end{document}